\documentstyle[aps,epsfig,prl,floats]{revtex}  
\newcommand{\scbo}{SrCu$_2$(BO$_3$)$_2$ }

\newcommand{\be}{\begin{equation}}
\newcommand{\ee}{\end{equation}}
\newcommand{\bea}{\begin{eqnarray}}
\newcommand{\eea}{\end{eqnarray}}

\begin{document}

\draft

\title{\bf The Phase Diagram of the Shastry-Sutherland Antiferromagnet}
\author{Zheng Weihong\cite{byline1}, J. Oitmaa\cite{byline3} and  C.J. Hamer\cite{byline2}} 
\address{School of Physics,                                              
The University of New South Wales,                                   
Sydney, NSW 2052, Australia.}                      

\date{\today}

\maketitle 

\begin{abstract}
The Shastry-Sutherland model, which provides a representation of the
magnetic properties of \scbo, has been studied by series expansion methods
at $T=0$. Our results support  the existence of an
intermediate phase between the N\'eel long-range ordered phase and
the short-range dimer phase. They provide strong evidence against
the existence of helical order in the intermediate phase, and 
 and somewhat less strong evidence against a plaquette-singlet phase.
 The nature of the intermediate phase thus remains elusive.
 It appears to be gapless, or nearly so.
\end{abstract}                                                              
\pacs{PACS numbers:  75.10.Jm., 75.40.Gb  }


\section{INTRODUCTION}

The recent discovery of the two-dimensional spin gap system 
\scbo\cite{kag98}, and its representation as an antiferromagnetic 
$S=\case 1/2$ Heisenberg spin model\cite{miy98} equivalent to
a model previously introduced by Shastry and Sutherland\cite{shastry},
has led to many recent studies of this model.

\begin{figure}[ht] 
\vspace{0cm}
\centerline{\hbox{\psfig{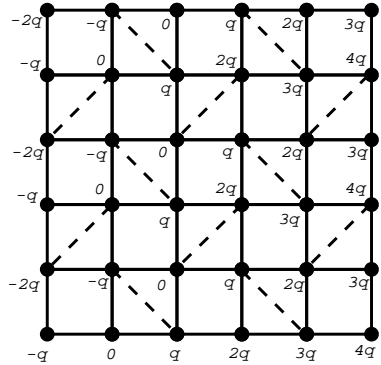}}}
\par
\caption{The Shastry-Sutherland lattice. The exchange $J'$
acts between sites separated by horizontal and vertical links, while
the exchange $J$ acts across the diagonal dashed links. The spin orientation 
at each site for
helical order is also given (near the sites), where $q=\arccos (-J'/J)$.
}
\label{fig_1}
\end{figure}

The Shastry-Sutherland model is a nearest-neighbor square lattice antiferromagnet,
with additional diagonal interactions, in a staggered pattern, on alternate
squares. The Hamiltonian is written as

\begin{equation}
H = J \sum_{{\rm diag}} {\bf S}_i\cdot {\bf S}_j 
+ J' \sum_{{\rm axial}} {\bf S}_i\cdot {\bf S}_j~, \label{H}
\end{equation}
This convention is adopted because in \scbo the diagonal bonds are,
in fact, the shortest and the exchange constants are estimated to be
$J \simeq 100K$, $J'\simeq 68K$.
The model is illustrated in Figure \ref{fig_1}.

It is clear that the system will exhibit N\'eel order for $J/J'$ small,
and will form a gapped spin-singlet state for $J/J'$ large, with $S=0$
dimers on the diagonal bonds. Shastry and Sutherland showed that this dimer
state is an exact eigenstate of the Hamiltonian for any $J, J'$ and that it is
rigorously the ground state for $J/J'>2$\cite{shastry}.

The main controversial and challenging  question  is whether the system has
an intermediate phase, and what kind of phase it is, if it exists.
In the classical limit, when $S\to \infty$, the ground state is N\'eel 
ordered if $J/J'\leq 1$ and is helically ordered (see Fig. \ref{fig_1}) otherwise, where the
twist between one spin and its nearest neighbor is given by
$q=\arccos (-J'/J)$. For the $S=\case 1/2$ system, 
Albrecht and Mila\cite{alb96} used a Schwinger boson mean-field theory
(SBMFT) and found an intermediate
phase with helical long-range order (LRO) for $1.1<J/J'<1.65$, but with $q$ differing 
from its classical value. 
In our previous series study\cite{us}, 
we were unaware of this  work, 
and  did not consider the possibility of  a  helical 
LRO intermediate phase. We  computed series for the
 N\'eel-ordered phase and for the dimer  phase at $T=0$, and located a transition point at
$J/J'=1.45(1)$ which appeared to be first-order. 

Subsequently Koga and Kawakami\cite{kog00} calculated series expansions 
about disconnected plaquettes, and claimed to have identified  two transition points:
a second-order transition from N\'eel order to a plaquette singlet phase at
$J/J'\simeq 1.16$, followed by a first-order transition to the dimer phase at
$J/J'\simeq 1.48$. We find this interpretation unconvincing,
partly because of uncertainty in the analysis of rather short series, and
partly because such an intermediate phase is totally different from
the helical-ordered phase presented in Ref. \onlinecite{alb96}.
Another recent development is the field-theoretical study of
a generalized model with $Sp(2N)$ symmetry by Chung, Marston and Sachdev
\cite{chu01}.  For $S=\case 1/2$, they suggest a N\'eel to helical LRO
phase transition at $J/J'\simeq 1.02$, and a helical LRO to
dimer short-range ordered (SRO) phase transition at  $J/J'\simeq 2.7$. This is similar to the scenario 
in Ref. \onlinecite{alb96}, but with the helical phase extending over
a much larger range. The quantitative accuracy of this approach for
$S=\case 1/2$ is problematic. Both our previous work\cite{us} and
the plaquette series of Koga and Kawakami\cite{kog00} provide strong
evidence for the dimer phase setting in at around $J/J' \simeq 1.45 - 1.48$.
It is also interesting to note that for $1/S>5$ (S is a continuous variable
in this theory) a phase with plaquette SRO is predicted\cite{chu01}. If finite $N$ fluctuations
change the phase diagram significantly then it is conceivable,
according to this theory, that there are four stable phases.

Knetter {\it et al.}\cite{kne00} subsequently performed a new series
analysis of the dimer phase. They find that the lowest single-triplet excitation 
vanishes at $J'/J=0.69$ (or $J/J'=1.45$), very close to the transition point mentioned above; 
but that an $S=1$ two-triplet excitation energy vanishes even earlier,
at $J'/J=0.63$ (or $J/J'=1.59$). This would seem to indicate important binding between the 
triplets, possibly leading to a condensate of triplets at or even below
$J'/J=0.63$. Even more recently, Totsuka, Miyahara, and Ueda\cite{tot01}
have discussed two-triplet binding effects  in this model using
both perturbation theory and exact diagonalization, and
have shown that binding occurs even in the quintet channel of two
triplets.

All the above studies show that the quantum system is N\'eel ordered  up to and 
beyond $J/J'=1$, 
i.e. beyond the regime for the classical system: this is also consistent with the 
general phenomenon of quantum order by disorder\cite{cha90}.
The transition  to dimer order is located at 
$J/J' \simeq 1.45 - 1.58$, so if there is an intermediate phase, it must exist
within the region $ 1 < J/J' \lesssim 1.5 $.
Our aim, in the present paper, is to explore these issues through new extended series expansions.
We have obtained extended series expansions about isolated plaquettes, both
with and without diagonal bonds, following Koga and Kawakami\cite{kog00}.
Secondly, we have extended our previous series about the N\'eel
ordered state and the dimer singlet state\cite{us}. Finally, we present
new series expansions about states with helical LRO and columnar dimer order. 
The next Section presents technical details of the calculations, with analysis
of series and discussion of results given in Section III.
In the last section we present our conclusions.

We conclude that an intermediate phase between the N\'eel phase and
the dimer phase very likely exists, in the region $1.2\lesssim J/J' \lesssim 
1.5$. The nature of this phase is much less certain, however.
It does not appear to be plaquette ordered or helically ordered.
It could perhaps posses weak columnar dimer order; or it could posses some other type of
order, not considered here. 
In any case, this intermediate phase appears to be gapless, or nearly so.

\section{Series Expansions}

We use the linked-cluster expansion method to derive perturbation expansions
for various choices of the unperturbed reference Hamiltonian. The technical
details are discussed in many articles, and we refer the reader who is unfamiliar
with these to a recent review\cite{serreview}. The following is a brief summary,

\subsection{Dimer Expansions}
The first term in the Hamiltonian (\ref{H}), which represents disconnected dimers, is taken
as the unperturbed Hamiltonian. 
The unperturbed ground state is then a product state of $S=0$ 
singlets on these dimers. As mentioned above, this state remains an exact eigenstate of the 
system for all $J,J'$, but is not the true ground state 
for $ J'/J > (J'/J)_c $. The perturbation,
 being the second term in (\ref{H}), mixes this state with states in which triplet
 excitations occur on the dimers.

Dimer expansions can be developed for both ground state properties and for excitations.
Here because the dimer state is the exact ground state, we focus on the triplet excitation spectrum
$\Delta ( {\bf k} )$. We have computed expansions  to order $(J'/J)^{21}$, extending  
our previous calculation\cite{us} 
 by 6 terms. This
calculation involves 185332 linked clusters up to 11 sites.
The resulting series coefficients are available on request. The minimum gap is at 
${\bf k}=0$, where the series is:
\bea
\Delta ({\bf k}=0)/J &&  = 
  1 - {x^2} - {\frac{{x^3}}{2}} - {\frac{{x^4}}{8}} + {\frac{5\,{x^5}}{32}} - 
   {\frac{7\,{x^6}}{384}} - {\frac{2051\,{x^7}}{4608}} - {\frac{39091\,{x^8}}{55296}} \nonumber \\
   &&  - {\frac{268849\,{x^9}}{663552}} + 
   {\frac{964411\,{x^{10}}}{6635520}} + 
   {\frac{6597973\,{x^{11}}}{58982400}} - 
   {\frac{183919894867\,{x^{12}}}{191102976000}} - 2.180975296\,{x^{13}} \nonumber \\
   &&  - 
   1.938901500\,{x^{14}} - 
   0.071485040\,{x^{15}} + 
   0.820962395\,{x^{16}} - 
   2.236236507\,{x^{17}} \nonumber \\
   && - 7.855506946\,{x^{18}} - 
   9.452785582\,{x^{19}} - 
   2.833667089\,{x^{20}} + 
   4.387303805\,{x^{21}} + O(x^{22})
\eea
where $x=J'/J$.
A standard Dlog Pad\'e approximant analysis\cite{gut} gives the gap vanishing 
at 
$J'/J=0.6965(15)$, or $J/J'=1.436(3)$
with the estimated critical index $\nu \simeq 0.45(2)$. 
Different Dlog Pad\'e approximants show remarkable consistency
with this exponent value, suggesting that this transition may belong to a new
universality class. These results are in agreement with M\"uller-Hartmann {\it et al.}\cite{har00}, 
who used shorter series.
But we recall that Knetter {\it et al.}\cite{kne00} found that a two-triplet
$S=1$ bound-state engery vanishes even before $J'/J=0.70$,
at around 0.63, which would entirely alter the position and the nature of this
transition. We have not computed series for the two-particle states
in this model. It would be very interesting to explore this transition further,
using series or other methods.

We also computed a new series, 
to order $(J'/J)^{17}$,
for the susceptibility $\chi$ corresponding to this momentum
${\bf k}=0$. The resulting series is
\bea
\chi ({\bf k}=0) &&  = 
1 + \frac{x^2}{2} + \frac{x^3}{8} + \frac{41\,x^4}{96} + 
  \frac{281\,x^5}{1152} + \frac{6461\,x^6}{13824} + 
  \frac{65953\,x^7}{165888} + \frac{186863\,x^8}{311040} \nonumber \\      
&&  +  \frac{374542669\,x^9}{597196800} + 
  \frac{24573240371\,x^{10}}{28665446400} + 
  \frac{2138459511091\,x^{11}}{2149908480000} + 
  1.308076648\,x^{12} \nonumber \\  
  && +  1.604414343\,x^{13} + 
  2.086275141\,x^{14} + 
  2.638378225\,x^{15} + 
  3.431256777\,x^{16} + 
  4.419063264\,x^{17} + O(x^{18})
\eea
A standard Dlog Pad\'e analysis of this series indicates a preferred
critical point around $J'/J=0.71$. Biasing $x_c$ at
0.6965, the critical exponent is found to be very small, around 0.03,
which is completely different from  that of the classical 
3D Heisenberg model. It might even be compatible with a logarithmic 
divergence.

\subsection{Plaquette Expansions}
\begin{figure}[ht]
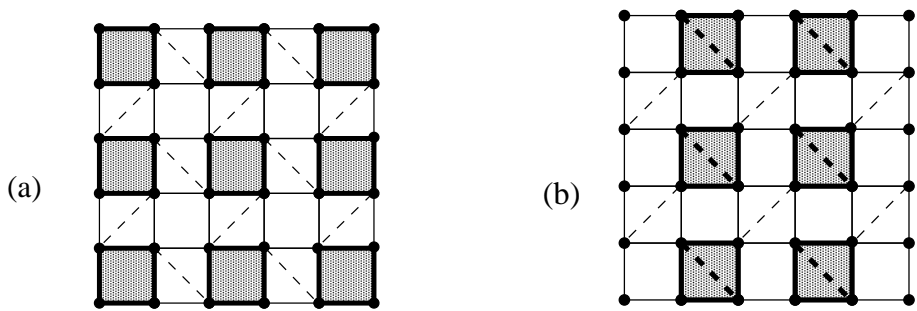
 
\vspace{0cm}
\centerline{\hbox{\psfig{figure=fig2a.eps,width=5cm}}\hspace{2cm} \hbox{\psfig{figure=fig2b.eps,width=5cm}}}
\par
\caption{(a) The first plaquette structure (PE1): The bold solid, the thin solid and 
the dashed lines represent the coupling constants $J'$, $\lambda J'$ and
$\lambda J$, respectively.  (b) The second plaquette structure (PE2): 
The bold solid, bold dashed, the thin solid, and the thin 
dashed lines represent the coupling constants $J'$, $J$, $\lambda J'$ and
$\lambda J$, respectively. 
}
\label{fig_2}
\end{figure}

Instead of perturbing about isolated dimers, any set of disconnected units can be used.
Using a plaquette basis allows the investigation of plaquette type order.
Because each plaquette has 16 rather than 4 states it is not possible to derive
series of the same length as with dimer expansions.

Following Koga and Kawakami\cite{kog00}, we have computed two kinds of plaquette expansions (PE1 and PE2),
which respectively take the set of plaquettes without the diagonal $J$ bonds
and the set with diagonal $J$ bonds as the unperturbed Hamiltonian.
To make the expansion possible it is necessary to introduce
 an  expansion parameter $\lambda\leq 1$ which modifies the interactions not included in $H_0$.
 This is illustrated in Figure 2. The series are then computed in powers of
 $\lambda$. The analysis evaluates these at $\lambda=1$, corresponding to the
 original Hamiltonian.
 
For PE1 we have computed series to order $\lambda^8$, $\lambda^7$, $\lambda^6$ for the ground state
energy, triplet excitation energies $\Delta ({\bf k})$
 and staggered susceptibility $\chi_{\rm AF}$
respectively. This is one additional term for $E_0$ and two additional terms for 
$\Delta ({\bf k})$ and $\chi_{\rm AF}$ over Ref. \onlinecite{kog00}.
The calculation is computationally demanding, even with our efficient program.
For example, the computation of $\Delta ({\bf k})$ to order $\lambda^7$ took about 10 days
and required 1.8GB memory on an SGI Origin 2400 system with a 400MHz R12000 CPU.
The effort for each additional term requires factors of approximate 50 and 15 increase in CPU
time and memory, respectively. For the second expansion (PE2) the series have been computed to order $\lambda^7$
for the ground state energy and the singlet excitation spectrum, and to order
$\lambda^6$ for the triplet excitation $\Delta ({\bf k})$ 
and staggered susceptibility $\chi_{\rm AF}$. In Tables \ref{tab1} and \ref{tab3} we present
the series for the 
 ground state energy, the triplet gap $\Delta$ at ${\bf k}=0$ and  staggered susceptibility
$\chi_{AF}$ for $J/J'=1.25$ and 1.4.
Other series are available on request. The analysis is left for the next Section.

\subsection{Columnar Dimer Expansions}

\begin{figure}[ht] 
\vspace{0cm}
\centerline{\psfig{figure=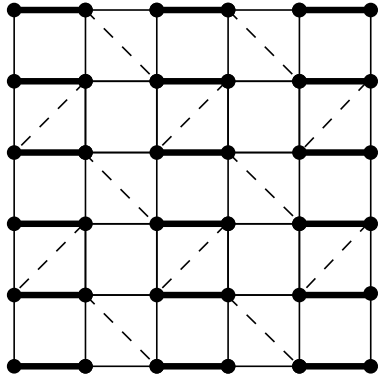,width=5cm}}
\par
\caption{
Columnar dimerization pattern. The bold 
solid, the thin solid and 
the dashed lines represent the coupling constants $J'$, $\lambda J'$ and
$\lambda J$, respectively.  
}
\label{fig_3}
\end{figure}

A plausible candidate for the
 intermediate phase is
 the columnar dimer phase, as shown in Fig. 3.
 For both the spin-${1\over 2}$ 
$J_1-J_2$ Heisenberg model\cite{raj99} 
and   the spin-${1\over 2}$ Heisenberg model on 
an anisotropic triangular lattice\cite{zwh99},
there appears to be a columnar dimer ordered phase 
adjacent to the N\'eel ordered phase.
To make the expansion possible it is necessary to introduce
 an  expansion parameter $\lambda\leq 1$ which modifies the 
interactions not included in $H_0$.
 This is illustrated in Figure \ref{fig_3}. The series are 
 then computed in powers of
 $\lambda$. The analysis evaluates these at $\lambda=1$, corresponding to the
 original Hamiltonian.
 
We have computed series to order $\lambda^{10}$, $\lambda^9$, $\lambda^9$ 
for the ground state
energy, the antiferromagnetic susceptibility $\chi_{\rm AF}$,
and the triplet excitation energies $\Delta ({\bf k})$
respectively. This
calculation involves 290215 linked clusters up to 10 sites.
In Table \ref{tab1}  we present
the series for the 
 ground state energy 
$\chi_{AF}$ for $J/J'=1.25$ and 1.4.
Other series are available on request. The analysis is left for the next Section.

\subsection{Ising Expansions for N\'eel order }
For small  $J/J'$, the model will have a  N\'eel ordered phase 
with long range antiferromagnetic order (in the $z$ direction).
To construct an Ising expansion, we introduce an exchange anisotropy parameter $\lambda$,
and write the Hamiltonian  as
\begin{equation}
H = H_0 + \lambda V~,  \label{Hising}
\end{equation}
where
\begin{eqnarray}
H_0 &= &  J \sum_{{\rm diag}} S_{i}^z S_{j}^z
+ J' \sum_{{\rm axial}} S_{i}^z S_{j}^z  ~, \nonumber \\
V &= J & \sum_{{\rm diag}}
 ( S_{i}^x S_{j}^x + S_{i}^y S_{j}^y )  +
J' \sum_{{\rm axial}} ( S_{i}^x S_{j}^x + S_{i}^y S_{j}^y )~,
\end{eqnarray}
The unperturbed  ground state is the classical N\'eel state with energy 
$E_0/N= -J'/2+J/8$. The perturbation $V$ flips pairs of neighbouring spins.
Series are developed in powers of $\lambda$ and evaluated at $\lambda=1$, which recovers
the original Hamiltonian. We have obtained series for
the ground state energy per site, $E_0/N$, and
the staggered magnetization (order parameter) $M$ to order $\lambda^{12}$,  
 extending  our previous calculation 
 by 3 terms. We have also computed a new series, to order $\lambda^{12}$ for the perpendicular
 susceptibility $\chi_{\perp}$.
The resulting series for the ground state energy and 
the staggered magnetization (order parameter) $M$
for $J/J'=1.25$ and 1.4
 are listed in
 Tables \ref{tab1} and \ref{tab2};  the series for other values of $J/J'$ 
 are
available on request. For details of the analysis we refer to our previous paper\cite{us}.
Results of the analysis are presented in the next section.

\subsection{Ising Expansions for Helical Order}

As discussed by Albrecht and Mila\cite{alb96}, the classical system
has planar  helical order for $J/J'>1$.
Starting from a reference spin in the $z$ direction, each neighboring spin is
rotated by an angle $q$, as shown in Figure 1. The twist is determined by 
 minimization of  the energy and yields
\begin{equation}
q = \cases {\pi, & $J\le J'$;\cr
\arccos (-J'/J), &$J > J'$.\cr} \label{eq_q}
\end{equation}

To develop an Ising expansion about such a helically ordered state we
transform the Hamiltonian by rotating the spin axes at each site. The transformed
Hamiltonian is
\begin{equation}
H = H_1 + \lambda ( H_2 +  H_3 ) ,  \label{Hq}
\end{equation}
where
\begin{eqnarray}
H_1 &=&  J \cos (2q) \sum_{\rm diag} S_{i}^z S_{n}^z
+ J' \cos (q) \sum_{\rm axial } S_{i}^z S_{j}^z  ~, \nonumber \\
H_2 &=& J \sum_{\rm diag} \left[ S_{i}^y S_{n}^y + \cos(2q) S_{i}^x S_{n}^x 
   + \sin (2q) ( S_{i}^z S_{n}^x - S_{i}^x S_{n}^z ) \right] ~, \label{Hq012} \\
H_3 &=& J' \sum_{\rm axial} \left[ S_{i}^y S_{j}^y + \cos(q) S_{i}^x S_{j}^x 
   + \sin (q) ( S_{i}^z S_{j}^x - S_{i}^x S_{j}^z ) \right] ~. \nonumber 
\end{eqnarray}
and where exchange anisotropy is introduced through the perturbation
parameter $\lambda$.

We have computed series for the ground state energy and the order parameter
to order $\lambda^{12}$, for various choices of $q$ and the ratio $J/J'$.
The resulting series for the ground state energy and 
the staggered magnetization (order parameter) $M$
for $J/J'=1.25$ and 1.4
 are listed in
 Tables \ref{tab1} and \ref{tab2};  other series 
 can be supplied on request. We are not aware of previous series of this
type for this model.

\section{Results and Discussion}
In this Section we present a variety of results obtained from analysis of the
various series expansions. The analysis  has been carried out using
integrated first-order inhomogeneous
differential approximants and  Pad\'{e} approximants\cite{gut}
to extrapolate each series to the {\it physical} value $\lambda=1$.


\begin{figure}[ht] 
\vspace{0cm}
\centerline{\hbox{\psfig{figure=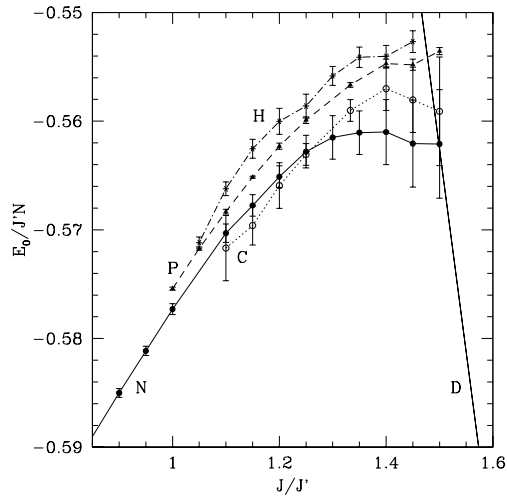,width=8cm}}}
\par
\caption{The ground state energy versus $J/J'$ obtained from  the N\'eel ordered Ising
expansion (N), columnar dimer expansion (C), plaquette expansion (P), and 
helical order Ising expansion (H). The energy of the exact dimer 
state is also presented (D). 
}
\label{fig_E0_y}
\end{figure}

\subsection{Ground State Energy}

The nature of the ground state for any particular $J/J'$ can be determined by
comparing the energies of various candidate states.
Our analysis shows that, among all the series expansions we have calculated, the 
ground state energy from the Ising expansion about N\'eel order has the lowest
energy for coupling $J/J' < 1.45$, while
the ground state energies obtained from the columnar dimer expansion and
from the N\'eel order Ising expansion
remain nearly equal through $J/J' \gtrsim 1.2$.
For example, we estimate the ground state energy
at $J/J'=1.25$ as
\be
E_0/NJ' = \cases 
{-0.5628(15), & N\'eel order;\cr
-0.5586(10), & helical order; \cr
-0.5599(3), & PE1; \cr
-0.5584(8), & PE2; \cr
-0.563(1), & columnar dimer. \cr
}
\ee
and at $J/J'=1.40$ as
\be
E_0/NJ' = \cases 
{-0.561(3), & N\'eel order;\cr
-0.554(1), & helical order; \cr
-0.5548(5), & PE1; \cr
-0.5525(10), & PE2; \cr
-0.557(2), & columnar dimer. \cr
} \label{eq_E0_140}
\ee
where we have used the classical-$q$ value in Eq. (\ref{eq_q}) in the calculation of
helical order.
In Table \ref{tab_e0_ida} we show details of the analysis from the latter case
 $J/J'=1.4$ for both the Ising expansion about N\'eel order  
 and the plaquette expansion (PE1), from which we
 estimate the values in (\ref{eq_E0_140}). 
The ground state energies obtained from various expansions 
versus $J/J'$ are given in Fig. \ref{fig_E0_y}. Since the energy from
the second plaquette expansion is slightly higher than that from the first 
plaquette expansion, we do not show the results of the
second plaquette expansion.
Our results are thus in disagreement with
ref. \onlinecite{kog00}, which claimed that the plaquette phase has the lowest energy.
Of course these energies are all very close to each other and the estimated errors are
subjective confidence limits. However inspection of the data in Table \ref{tab_e0_ida}
shows that the Ising expansion about N\'eel order  consistently gives lower
energy estimates.

\begin{figure}[ht] 
\vspace{0cm}
\centerline{\hbox{\psfig{figure=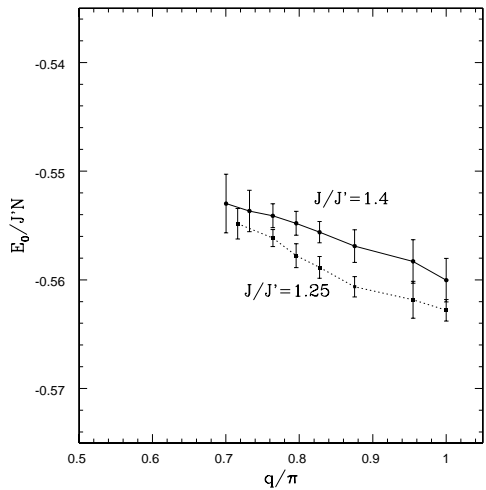,width=8cm}}}
\par
\caption{The ground state energy versus $q$ for helical order. 
}
\label{fig_E0_helical}
\end{figure}

 Our results also argue against helical order being a stable ground state.
 In Figure \ref{fig_E0_helical} we show estimates of the ground state energy 
 from Ising expansions about helical order, as a function of angle $q$,
 for the two coupling ratios $J/J'=1.25$, 1.40. In each case
 the curve shows no indication of a minimum at some $q_c$, but rather
 decreases monotonically towards
  $q=\pi$, corresponding to N\'eel order.
 
Over the region $1.2\lesssim J/J' \lesssim 1.45$, the ground state energies
obtained from the N\'eel ordered Ising expansion and from the columnar dimer expansion
have some overlap after considering the error bars, so they are both
 good candidates for the ground state of the Shastry-Sutherland model.

\subsection{Staggered Magnetization and Perpendicular Susceptibility}

\begin{figure}[ht] 
\vspace{0cm}
\centerline{\hbox{\psfig{figure=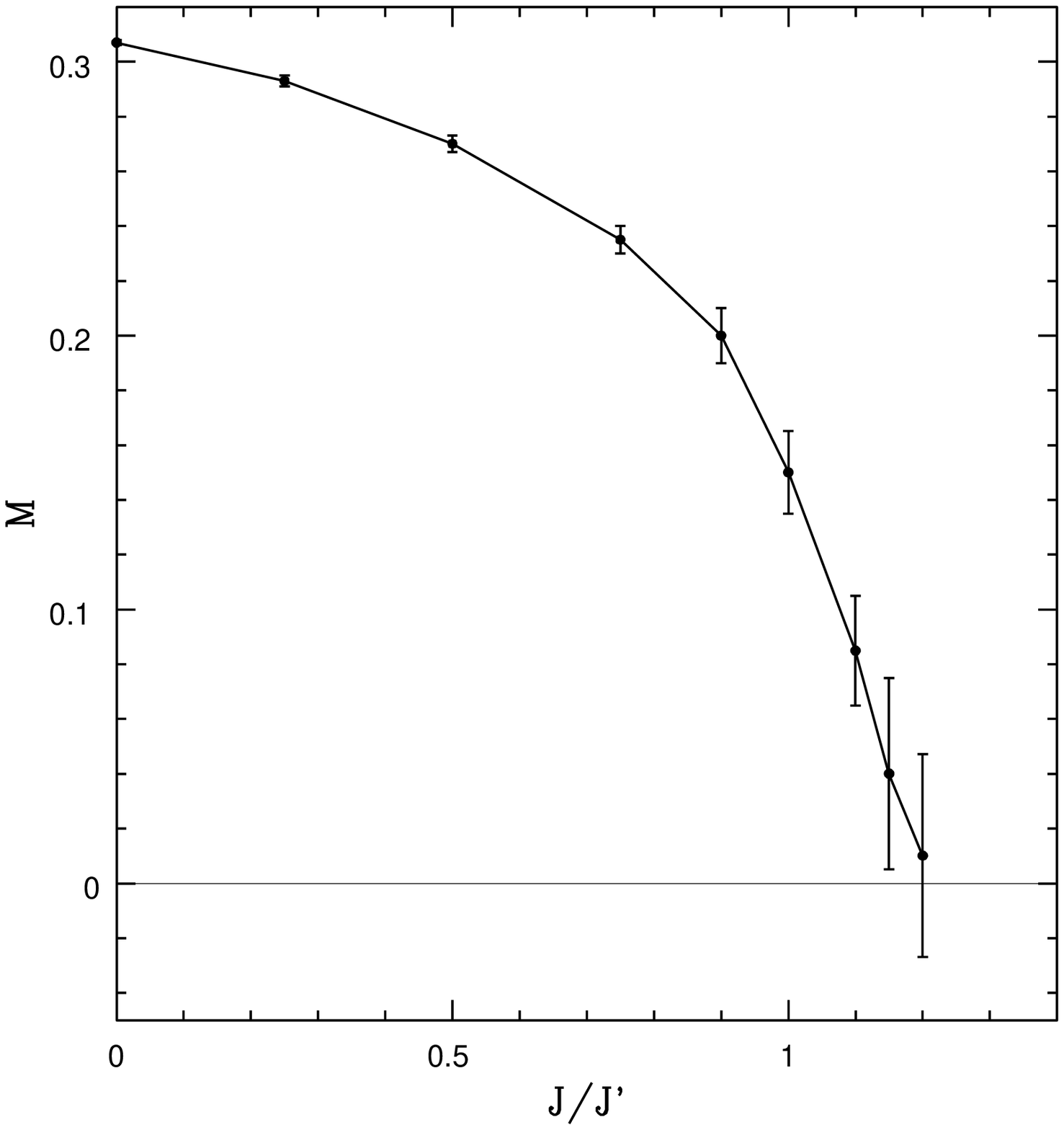,width=7cm}} \hspace{0cm} \hbox{\psfig{figure=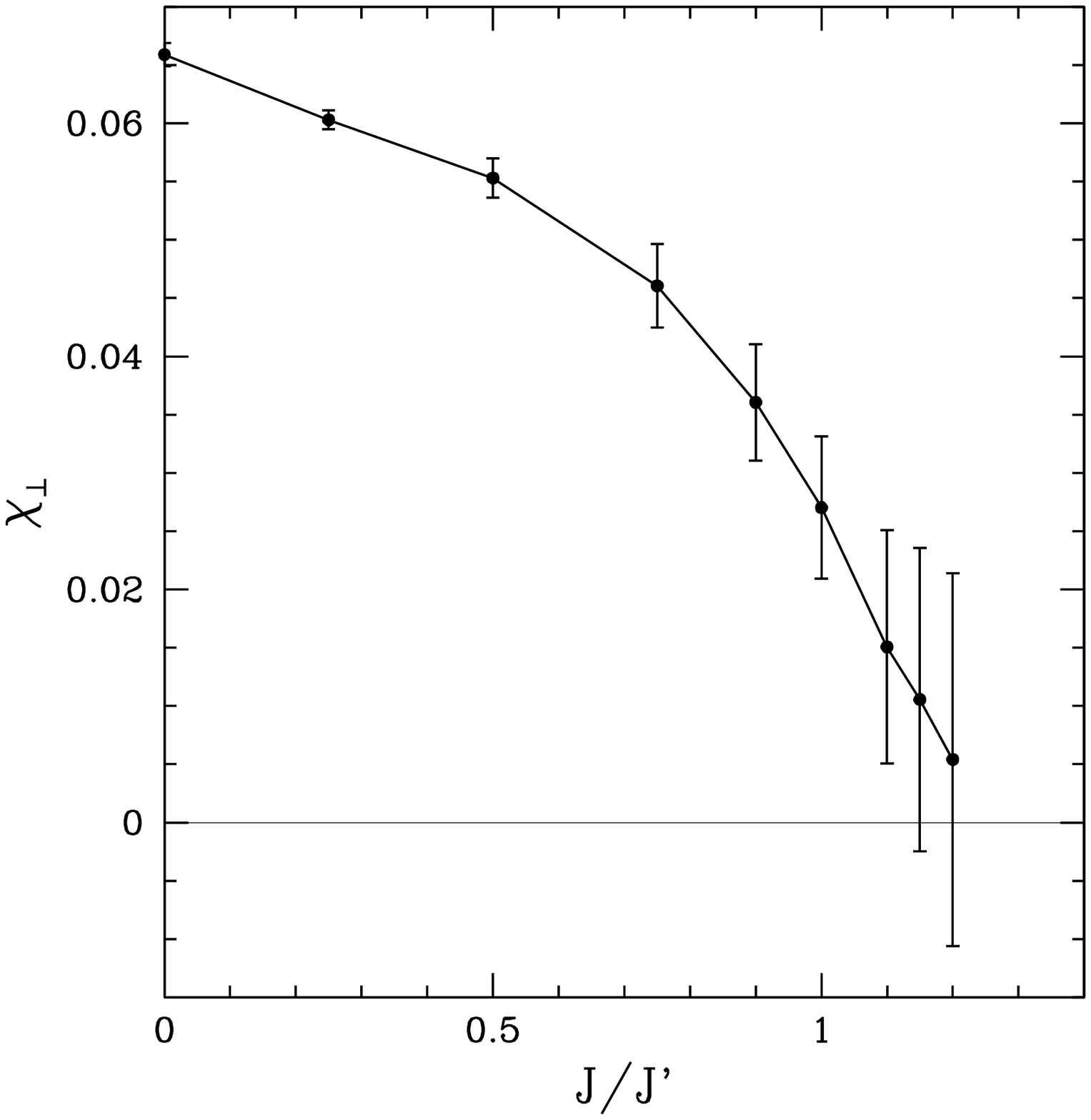,width=7cm}}}
\par
\caption{The staggered magnetization $M$  and  the perpendicular susceptibility $\chi_{\perp}$
{\it versus} $J/J'$.
The solid points with error bars are the estimates from
the Ising expansion about N\'eel order.  
}
\label{fig_M_chi}
\end{figure}

The staggered magnetization and perpendicular susceptibility will be non-zero
in a phase with long range antiferromagnetic order, and are expected to
vanish at a transition point to a magnetically disordered or spin-liquid
phase. Effective Lagrangian theory predicts a relationship
$\rho_s=v^2 \chi_{\perp}$ where $\rho_s$ is the spin stiffness and $v$ the
spin-wave velocity. In Figures \ref{fig_M_chi}  we show estimates of $M$ and
$\chi_{\perp}$ versus $J/J'$, from the Ising expansion about
N\'eel order. It can be seen that both quantities behave similarly, decreasing from
their values at $J=0$ and vanishing at around $J/J'=1.2\pm 0.1$.
The error bars are large and the vanishing point can not be obtained to high precision.
However our new extended series shows a much sharper drop-off in $M$ than given in our 
previous work\cite{us}. Thus it is possible that the true transition point is at or below
1.20.

For the Ising expansion about helical order the magnetization $M$ at $\lambda=1$ (for the classical $q$ value)
is zero, within error bars, for all $J/J'$. This strengthens our conclusion, 
from the ground state energy results, that a helical phase is not present.

\subsection{Energy gap and susceptibility in the plaquette expansion}

The main argument of Koga and Kawakami\cite{kog00}  for the
existence of a plaquette intermediate phase in the system is based 
on analysis of the triplet gap at ${\bf k}=0$
and the staggered susceptibility series. 
In their series analysis, they assumed that the minimum triplet gap is at ${\bf k}=0$, and
the transition  lies in the same universality class as the classical 3D Heisenberg model
(i.e. critical exponents $\gamma\simeq 1.4$, $\nu\simeq 0.71$).
Using standard Dlog Pad\'e approximants,
they found an apparent critical singularity at $\lambda>1$ for  $J/J' \geq 1.16$.
This would imply a non-zero spin gap
for $J/J' \geq 1.16$ for the original Hamiltonian (\ref{H}).
The above assumption is valid for the transition to N\'eel order, but if
there was an intermediate phase, the transition from the plaquette
phase to this intermediate phase might not lie in the same universality 
class as the classical 3D Heisenberg model, and  the minimum triplet gap 
might not be at ${\bf k}=0$.
In the analysis with our longer series, we find indeed  
that the minimum triplet gap is not at ${\bf k}=0$  for $J/J' \gtrsim 1.25$,
and also that the transition does not lie 
in the classical 3D Heisenberg  universality class.

\begin{figure}[ht] 
\vspace{0cm}
\centerline{\psfig{figure=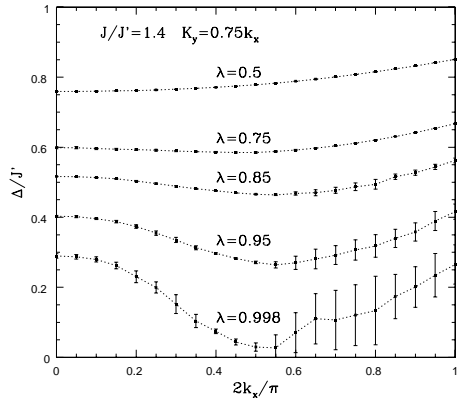,width=7cm}} 
\par
\caption{The triplet dispersion for various $\lambda$ and $J/J'=1.40$
along the line connecting ${\bf k}=(0,0)$
and ${\bf k}=(\pi/4,3 \pi/16 )$ obtained from 
the first plaquette expansion (PE1).
}
\label{fig_mk_1p40}
\end{figure}

Firstly, let us discuss the first plaquette expansion (PE1).
Here  for $J/J'\lesssim 1.25$, or for larger values of $J/J'$ with 
 small values of $\lambda$, 
we find the minimum triplet gap is at
${\bf k}=0$; but for $J/J'\gtrsim 1.25$ and $\lambda\sim 1$, 
it is no longer at
${\bf k}=0$. For example for $J/J'=1.40$, the dispersion for various $\lambda$ 
along the line connecting ${\bf k}=(0,0)$
and ${\bf k}=(\pi/4,3 \pi/16 )$ is shown in Fig. \ref{fig_mk_1p40}, where
we can see  that for $\lambda\lesssim 0.75$, the minimum gap is located at
${\bf k}=0$, while for  larger $\lambda$, the minimum gap 
is located at about ${\bf k}=(\pi/4, 3 \pi/16)$, and the 
dispersion near the  minimum is quite flat. 
The series for the triplet gap at ${\bf k}=(\pi/4, 3 \pi/16)$ is
\bea
\Delta(\pi/4, 3 \pi/16 )/J' && =
   1      
 -0.3077152787          \lambda
 -0.2293662299          \lambda^2
 -0.0346064996          \lambda^3 \nonumber \\
&& -0.0741227651        \lambda^4 
 -0.0176489659          \lambda^5
 -0.0400694385          \lambda^6
 -0.0030161628          \lambda^7 + O(\lambda^8 )
\eea

Table V shows estimates of the critical point $\lambda_c$ and
critical index from both the spin  gap and susceptibility series for
the coupling ratios $J/J'=1.25$ and 1.40, using unbiased Dlog Pad\'e
approximants, where for $J/J'=1.40$, we show the results for the spin gap
at both ${\bf k}=0$ and ${\bf k}=(\pi/4, 3\pi/16)$.
Apart from the spin gap $\Delta(\pi/4, 3\pi/16)$ at $J/J'=1.40$,  
there is considerable scatter in the results, but several
features are apparent. The spin gap series show decreasing estimates of 
$\lambda_c$ with increasing order for both $J/J'$ ratios. The data are more
consistent with the conclusion $\lambda_c\leq 1$, rather than $\lambda_c>1$.
For the spin gap $\Delta(\pi/4, 3\pi/16)$ at $J/J'=1.40$, 
the unbiased Dlog Pad\'e
approximants give very convergent results, with the 
critical point $\lambda_c=0.988$
and critical index $\nu=0.43$. This 
 indicates instability of the plaquette phase at 
 $\lambda=1$, at least for this  coupling ratio.
 This also indicates that there is indeed an intermediate phase.
Note that the critical index $\nu$ obtained here is about
the same as that obtained from the dimer series,
suggesting that this transition might belong to the same new
universality class. 
Our analysis shows that 
the transition probably does not belong in the classical 3D Heisenberg universality 
class  for $J/J'\gtrsim 1.1$.
Because the critical index $\nu$ for the 
minimum gap 
is much smaller than 1, we have obtained the  dispersion
results shown in Fig. \ref{fig_mk_1p40} by performing the series
extrapolation in a new variable
\be
\delta = 1 - (1-\lambda/\lambda_c)^{\nu}~.\label{eq_delta}
\ee
in order to take the singular behavior into account.



Finally we consider the second plaquette expansion (PE2). The minimum triplet gap is no longer at
${\bf k}=0$, at least for small to moderate $\lambda$. This can be seen from the
first few terms of the series
\be
\Delta(k_x, k_y)/J' = 1 - {4\lambda \over 6} (\cos k_x + \cos k_y ) + {J \lambda \over 3 J' } 
\cos (k_x + k_y)
\ee
which gives the minimum gap at $k_x=k_y=\arccos (J'/J)$.
It seems likely that the minimum triplet gap will remain at ${\bf k}\not= 0$, even for $\lambda\simeq 1$,
although the series in this region are too erratic to confirm this.
For this choice of unperturbed Hamiltonian we are also
able to compute the excitation spectrum for the singlet excitation and we
have obtained series to order $\lambda^7$. The minimum singlet gap is at
${\bf k}=0$ and appears to be smaller than the triplet gap.
An attempt to locate the critical point $\lambda_c$ by Dlog Pad\'e approximants
to this series was hampered by poor convergence.
We have attempted other analysis procedures, but without great success.
But since the ground state energy from the second plaquette expansion (PE2)
is higher than that obtained from both the first plaquette expansion
and the N\'eel ordered Ising expansion, the probability of having the
second plaquette configuration as the intermediate phase is remote.

\begin{figure}[ht] 
\vspace{0cm}
\centerline{\hbox{\psfig{figure=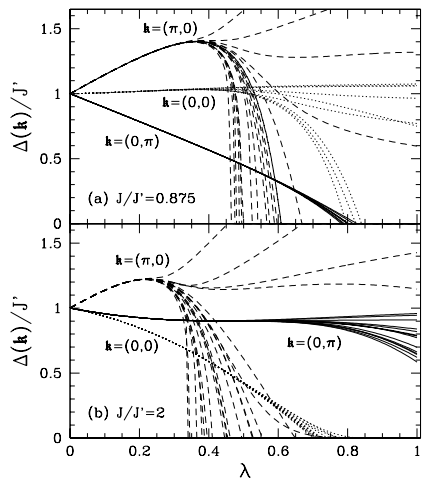,width=7cm}} }
\par
\caption{The triplet gap $\Delta({\bf k})/J'$ 
{\it versus} $\lambda$ for ${\bf k}=(0,0)$, $(\pi,0)$ and $(0,\pi)$ and $J/J'=0.875$, and 2.
Several different integrated differential
approximants to the series are shown.
}
\label{fig_gap_columnar_dimer}
\end{figure}

\subsection{The energy gap  in the columnar dimer expansion}

\begin{figure}[ht] 
\vspace{0cm}
\centerline{\hbox{\psfig{figure=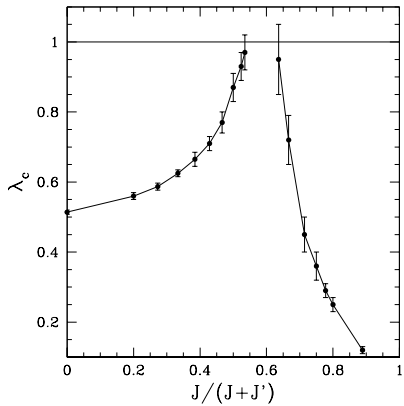,width=7cm}} }
\par
\caption{The phase diagram for columnar dimer order, where the curve in the small $J/(J+J')$ region
is determined by the Dlog Pad\'e approximants to the gap  at ${\bf k}=(0,\pi )$
and the antiferromagnetic
susceptibility $\chi_{AF}$, and we assume the transition  lies in 
at the same universality class as the classical 3D Heisenberg model.
The curve in the large $J/(J+J')$ region
is determined by the Dlog Pad\'e approximants to the gap  at ${\bf k}=(0,0 )$.
}
\label{fig_pha_diag_columnar_dimer}
\end{figure}

Finally we discuss the triplet dispersion obtained from the columnar 
dimer expansion. Here for  $J/J'\lesssim 1.15$, 
we find  the minimum gap is located at
momentum ${\bf k}=(0, \pi)$, as expected, since we
expect to have a transition to the N\'eel ordered phase for
small $J/J'$. Fig. \ref{fig_gap_columnar_dimer}
show the gap at ${\bf k}=(0,0)$, $(\pi,0)$ and $(0,\pi)$
for $J/J'=0.875$ versus $\lambda$ obtained from the integrated differential
approximants\cite{gut} to the series. We can see that
the minimum gap vanishes at a critical
value  $\lambda_c<1$. We also expect this transition (to Ne\'el order) 
to lie in the same universality class as the classical 3D Heisenberg model.
To determine the phase boundary of columnar dimer order,
we can use Dlog Pad\'e approximants to 
both the series for the minimum gap and for the
 antiferromagnetic susceptibility. To get more
reliable estimates for the critical point, we  assume  the
critical exponents to be $\gamma\simeq 1.4$, $\nu\simeq 0.71$.
The results are shown in the phase diagram  Fig. \ref{fig_pha_diag_columnar_dimer},
where we can see that $\lambda_c=0.514(6)$ for $J=0$, and 
that $\lambda_c$ increases for increasing
$J$. For $J/J'=1.15$, we estimate 
$\lambda_c=0.97(5)$, but unfortunately the biased Dlog Pad\'e
 results here are not accurate enough
to tell  whether $\lambda_c$ is larger than 1.

For $J/J'\gtrsim 1.75$, we find that the minimum gap is located at
${\bf k}=(0,0)$. Fig. \ref{fig_gap_columnar_dimer}
also shows the gap at ${\bf k}=(0,0)$, $(\pi,0)$ and $(0,\pi)$
for $J/J'=2$ versus $\lambda$.
Once again, one can locate the phase boundary by
using the Dlog Pad\'e approximants to 
the series for the  minimum gap, and the results are also shown in 
Fig. \ref{fig_pha_diag_columnar_dimer}. We see
that as $J/J'$ decreases, $\lambda_c$ increases, but again, 
the analysis is not accurate enough to tell whether
 $\lambda_c$ is larger than 1 in the intermediate region.

In the analysis, we also note that the gap at ${\bf k}=(\pi,0)$ shows some
peculiar features, as shown in Fig. \ref{fig_gap_columnar_dimer}. For all values
of $J/J'$, the majority of integrated differential approximants to the series
 show that
the gap increases for small $\lambda$, then decreases dramatically
to zero for $\lambda\sim 0.5$. The majority of Dlog Pad\'e approximants
 to the series
also show that the series has a critical point at $\lambda\sim 0.5$
with very small critical index ($\sim 0.2$). But we believe this peculiar behavious
is an artefact of the short series.

\begin{figure}[ht] 
\vspace{0cm}
\centerline{\hbox{\psfig{figure=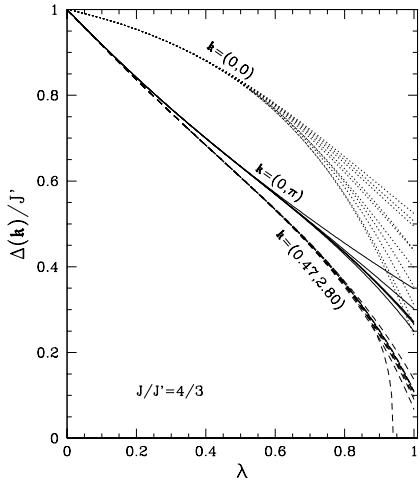,width=7cm}} }
\par
\caption{The triplet gap $\Delta({\bf k})/J'$ 
{\it versus} $\lambda$ for ${\bf k}=(0,0)$, $(0.47,2.80)$ and $(0,\pi)$ and $J/J'=4/3$.
Several different integrated differential
approximants to the series are shown.
}
\label{fig_gap_1p33}
\end{figure}

For the most interesting region $1.15  \lesssim J/J' \lesssim 1.75$,
the situation is quite complicated: the location of the minimum gap
depends on both $J/J'$ and  $\lambda$. We take the mid-point
for the presumed intermediate phase,  $J/J'=4/3$, as example.
For $\lambda$ near 1, we find the minimum gap for $J/J'=4/3$ is at
${\bf k}=(0.47,2.80)$, slightly away from $(0,\pi)$.
The standard Dlog Pad\'e approximants to the series for the minimum gap gives a 
critical point 
$\lambda_c=1.0(1)$, with critical index $\nu \sim 0.4$.
This seems to be consistent with the indices  
obtained from the dimer and plaquette expansions,
Since the standard Dlog Pad\'e approximants 
cannot tell whether $\lambda_c$ is larger than 1, we examine the
extrapolations for the gap itself  obtained from  integrated differential approximants.
The results 
for the gap at ${\bf k}=(0.47,2.80)$, $(0,0)$ and
$(0,\pi)$ are shown in Fig. \ref{fig_gap_1p33}.
The gap at $\lambda=1$ for ${\bf k}=(0.47,2.80)$ is very small ($\sim 0.1J'$)
when estimated from the direct $\lambda$-series. When analysed in term of
variable $\delta$ (Eq. \ref{eq_delta}) the gap is compatible with zero.
If the gap vanishes at or before $\lambda=1$, the columnar dimer
phase would also be ruled out as intermediate phase for the 
Shastry-Sutherland model.

\section{Conclusions}

We have attempted to further elucidate the nature of the phase diagram of the
Shastry-Sutherland spin model at $T=0$, by series expansion methods.
We have significantly extended previous series, computed by
ourselves\cite{us} and others\cite{kog00}, and we have also derived
a number of new series. The analysis of the various series allows us
to draw some fairly firm conclusions, as well as others which
are more tentative.

The first question is whether an intermediate phase between the N\'eel-ordered
phase and the dimer phase exists at all. It seems rather clear that the singlet
dimer phase, with a simple singlet-product ground state, persists from
large $J/J'$ down to $J/J'\simeq 1.5-1.6$. In our previous work\cite{us}
we argued that there is a direct first-order transition from the dimer phase to N\'eel
order at this point. Our present results
show the staggered magnetization and perpendicular susceptibility in the
N\'eel phase vanishing at $J/J'\simeq 1.2$, with large uncertainty
$\pm 0.1$.
This appears more consistent with
 the N\'eel phase terminating at a second-order phase transition, so that 
there may indeed be an intermediate phase stretching over 
a very small range of coupling constants $1.20 \lesssim J/J' \lesssim  1.5$,
in agreement with the suggestion of Koga and Kawakami\cite{kog00}.

Carpentier and
Balents\cite{car01} have recently discussed an effective mean-field theory approach
to the generalized Shastry-Sutherland model. They argue that a direct continuous
transition from the N\'eel phase to the dimer phase is not possible.
If this argument is accepted, it means that the vanishing of the order
parameter and susceptibility in the N\'eel phase cannot be associated
with a second order transition directly to the dimer phase.
It follows that an intermediate phase must exist, over at least a small
range of couplings.

The next question is the nature of the transitions
from the N\'eel and dimer phases, respectively, into
the intermediate phase. The vanishing staggered magnetization and
perpendicular susceptibility in the N\'eel phase 
indicate a 2nd order transition; but we are
unable to estimate the critical exponents, since we do not
have explicit series coefficients for an expansion in $J/J'$.
The vanishing energy gap would indicate
that the transition from the dimer phase is also 2nd order; but
the result of Knetter {\it et al.}\cite{kne00}
showing that the two-particle bound-state energy vanishes before 
the single triplet gap indicates that more work needs to be done to 
characterize this transition also. It is interesting that one-particle
gaps appears to vanish with a similar exponent in the dimer expansion, the
plaquette expansion, and also the columnar dimer expansion.



A further question concerns the nature of the intermediate phase.
None of the suggested phases exhibit
a ground-state energy which is distinctly lower than the N\'eel energy in the
 relevant coupling regime. Our results do not support the suggestion\cite{kog00}
that the intermediate phase is plaquette ordered. 
The extrapolated ground-state energy from the N\'eel expansion consistently 
lies below that from the plaquette expansion; and the triplet gap from
the plaquette expansion appears to vanish at or before the
physical value $\lambda=1$ is reached, indicating instability in this phase.
Our longer series thus appear to contradict the conclusions of
Koga and Kawakami\cite{kog00}.
Equally, our results do not support the suggestion that the 
intermediate phase is helically ordered\cite{alb96,chu01}.
The extrapolated ground-state energy from the N\'eel expansion
consistently lies below that of the helically-ordered state,
for any value of the spin orientation $q$ other than $\pi$.
We note that the mean-field approach of Albrecht and  Mila\cite{alb96}
is strictly
only valid for $d>2$; while the $1/N$ expansion approach of
Chung,  Marston, and  Sachdev\cite{chu01} is of doubtful
validity for the spin-$\case 1/2$ case, and is certainly not
quantitatively accurate.

We have also explored the possibility that the
intermediate phase is a columnar dimer phase, as in the
$J_1-J_2$ Heisenberg model\cite{raj99}, or
the spin-${1\over 2}$ Heisenberg model on 
an anisotropic triangular lattice\cite{zwh99}. The extrapolated ground state
energy for this state is comparable with that of the N\'eel
state, within error bars, and the triplet gap from this expansion shows
a window $1.15\lesssim J/J' \lesssim 1.5$ in which it may remain finite, 
though very small, at the physical value $\lambda =1$. 
This remains a possible
candidate for the intermediate phase, although our data are not
sufficiently accurate to allow a definitive conclusion.

The nature of an intermediate state in this model thus remains an open question.
It could, for example,
be a structureless spin liquid, in which case a signature of this phase 
is elusive. 
Carpentier and
Balents\cite{car01} have also discussed possible intermediate phases in the generalized model, 
including a WISDQ (weakly incommensurate spin-density wave) ordered state -i.e.
a periodic modulation of the expectation value of the total spin on each dimer - and 
a `fractionalized' state, with topological order and deconfined spin-$\case 1/2$
excitations (`spinons').
 We currently have no information to offer regarding these states.
 The fact that the energy gap drops to zero before or near the physical
 value $\lambda =1$ in every expansion we have tried appears to
 suggest that the intermediate phase is gapless, or near to it.
 Further support for this idea is given by the strong finite-size
 dependence of the exact diagonalization results of
 Miyahara and  Ueda\cite{miy98} in this region.
 


The material \scbo appears to lie within the dimer phase of the Shastry-Sutherland
model. Original estimates $J/J' = 1.47$\cite{miy98,us} put
the material close to the phase transition point,
and it was argued that some of the unusual properties of the
material were due to this closeness. A more recent estimate\cite{kne00},
including interactions in the third direction, gives
$J/J' \simeq 1.66$, somewhat beyond even the point at which the
two-particle gap vanishes. This puts the material clearly within the dimer
phase and hence the question of intermediate phases is primarily
of theoretical interest.

It is clear that much work remains to be done on this 
fascinating model. It would be interesting to confirm and
extend the results of  Knetter {\it et al.}\cite{kne00}
concerning the transition from the dimer phase; and also
to try and construct series coefficients for  an
expansion in $J/J'$ in the N\'eel phase. New
numerical methods are badly needed to explore the
intermediate phase also: the best nmethod to employ here
remains a puzzle.

\acknowledgments
This work forms part of a research project supported by a grant 
from the Australian Research Council. 
The computation has been performed on an SGI origin 2400 
 computer. We are grateful for the computing resources provided
 by New South Wales Australian Centre for Advanced Computing and
 Communication (ac3) center.


\newpage
\widetext

\begin{table}
\squeezetable
\setdec 0.0000000000000
\caption{Series coefficients for the ground-state energy per 
site $E_0/(NJ')$ obtained from the 
Ising expansion about N\'eel-order (N\'eel), the helical-order (helical) expansion,
the columnar dimer expansion, 
and from the plaquette expansion  without diagonal bonds (PE1), 
and with diagonal bonds (PE2) for $J/J'=1.25,1.4$. Series coefficients of $\lambda^i$
 are listed.
}
 \label{tab1}
\begin{tabular}{r|rrrrr}
\multicolumn{1}{c|}{$i$} 
&\multicolumn{1}{c}{PE1}&\multicolumn{1}{c}{PE2} 
&\multicolumn{1}{c}{Helical} &\multicolumn{1}{c}{N\'eel} &\multicolumn{1}{c}{columnar dimer}  \\
\tableline
\multicolumn{1}{c|}{}&\multicolumn{5}{c}{$J/J'=1.25$} \\
  0 & \dec -5.000000000000$\times 10^{-1}$ & \dec -4.218750000000$\times 10^{-1}$ & \dec -3.562500000000$\times 10^{-1}$ & \dec -3.437500000000$\times 10^{-1}$ & \dec -3.750000000000$\times 10^{-1}$ \\
  1 & \dec  0.000000000000                 & \dec  0.000000000000                 & \dec  0.000000000000                 & \dec  0.000000000000                 & \dec  0.000000000000                 \\
  2 & \dec -3.575303819444$\times 10^{-2}$ & \dec -1.577687449269$\times 10^{-1}$ & \dec -2.054711318598$\times 10^{-1}$ & \dec -2.857142857143$\times 10^{-1}$ & \dec -1.904296875000$\times 10^{-1}$ \\
  3 & \dec -9.498878761574$\times 10^{-3}$ & \dec  2.403841825921$\times 10^{-2}$ & \dec  4.163537328971$\times 10^{-2}$ & \dec  1.020408163265$\times 10^{-1}$ & \dec  7.141113281250$\times 10^{-3}$ \\
  4 & \dec -6.520377148279$\times 10^{-3}$ & \dec -2.785619169517$\times 10^{-3}$ & \dec -4.669142343282$\times 10^{-2}$ & \dec -9.273368951135$\times 10^{-2}$ & \dec  9.905815124512$\times 10^{-3}$ \\
  5 & \dec -3.110192567684$\times 10^{-3}$ & \dec  1.333554456619$\times 10^{-3}$ & \dec  2.832044653274$\times 10^{-2}$ & \dec  1.575551772772$\times 10^{-1}$ & \dec -9.499887625376$\times 10^{-3}$ \\
  6 & \dec -2.051453756690$\times 10^{-3}$ & \dec -9.330263748650$\times 10^{-3}$ & \dec -3.363486912444$\times 10^{-2}$ & \dec -2.473894356829$\times 10^{-1}$ & \dec -1.780716847214$\times 10^{-2}$ \\
  7 & \dec -1.134481706160$\times 10^{-3}$ & \dec  9.793071057074$\times 10^{-3}$ & \dec  3.693336765114$\times 10^{-2}$ & \dec  4.097681140990$\times 10^{-1}$ & \dec  1.494069254463$\times 10^{-2}$ \\
  8 & \dec -7.196107350240$\times 10^{-4}$ &                                      & \dec -5.318671447970$\times 10^{-2}$ & \dec -8.162334082025$\times 10^{-1}$ & \dec -3.803820899197$\times 10^{-4}$ \\
  9 &                           &                                                 & \dec  7.567434418701$\times 10^{-2}$ & \dec  1.693646904376                 & \dec  9.729690462284$\times 10^{-4}$ \\
 10 &                           &                                                 & \dec -1.138657557136$\times 10^{-1}$ & \dec -3.494392524807                 & \dec -2.666137442909$\times 10^{-3}$ \\
 11 &                           &                                                 & \dec  1.726680824433$\times 10^{-1}$ & \dec  7.456679516027                 & \\
 12 &                           &                                                 &                                      & \dec -1.646408487950$\times 10^{1}$  & \\
\tableline                                                                                       
\multicolumn{1}{c|}{}&\multicolumn{5}{c}{$J/J'=1.40$}\\
  0 & \dec -5.000000000000$\times 10^{-1}$ & \dec -4.125000000000$\times 10^{-1}$ & \dec -3.535714285714$\times 10^{-1}$ & \dec -3.250000000000$\times 10^{-1}$ & \dec -3.750000000000$\times 10^{-1}$ \\
  1 & \dec  0.000000000000                 & \dec  0.000000000000                 & \dec  0.000000000000                 & \dec  0.000000000000                 & \dec  0.000000000000                 \\
  2 & \dec -3.392361111111$\times 10^{-2}$ & \dec -1.678565705128$\times 10^{-1}$ & \dec -1.943166023166$\times 10^{-1}$ & \dec -3.125000000000$\times 10^{-1}$ & \dec -1.950000000000$\times 10^{-1}$ \\
  3 & \dec -8.058449074074$\times 10^{-3}$ & \dec  2.952951931067$\times 10^{-2}$ & \dec  3.629865386622$\times 10^{-2}$ & \dec  1.367187500000$\times 10^{-1}$ & \dec  1.659375000000$\times 10^{-2}$ \\
  4 & \dec -5.431377864048$\times 10^{-3}$ & \dec -1.145691038736$\times 10^{-3}$ & \dec -4.218721381575$\times 10^{-2}$ & \dec -1.664227322346$\times 10^{-1}$ & \dec  1.704765625000$\times 10^{-2}$ \\
  5 & \dec -2.536510876037$\times 10^{-3}$ & \dec  2.724731917375$\times 10^{-3}$ & \dec  2.276170842636$\times 10^{-2}$ & \dec  3.302972722063$\times 10^{-1}$ & \dec -1.338582682292$\times 10^{-2}$ \\
  6 & \dec -1.754982806912$\times 10^{-3}$ & \dec -1.315792108890$\times 10^{-2}$ & \dec -3.202093475641$\times 10^{-2}$ & \dec -6.320111458811$\times 10^{-1}$ & \dec -1.870258387587$\times 10^{-2}$ \\
  7 & \dec -1.017310795905$\times 10^{-3}$ & \dec  1.282279331689$\times 10^{-2}$ & \dec  2.995390698796$\times 10^{-2}$ & \dec  1.313875979565                 & \dec  1.669764147147$\times 10^{-2}$ \\
  8 & \dec -7.033788209007$\times 10^{-4}$ &                                      & \dec -4.311390066885$\times 10^{-2}$ & \dec -3.190631154651                 & \dec -1.134220766188$\times 10^{-3}$ \\
  9 &                           &                                                 & \dec  5.834414925135$\times 10^{-2}$ & \dec  8.019181935184                 & \dec -6.547755059723$\times 10^{-3}$ \\
 10 &                           &                                                 & \dec -8.797286992572$\times 10^{-2}$ & \dec -2.044153224739$\times 10^{1}$  & \dec -1.637369775362$\times 10^{-3}$ \\
 11 &                           &                                                 & \dec  1.341535224738$\times 10^{-1}$ & \dec  5.383666306276$\times 10^{1}$  & \\
 12 &                           &                           &                                                            & \dec -1.457331428464$\times 10^{2}$  & \\
\end{tabular}                                                                             
\end{table}

\begin{table}
\squeezetable
\setdec 0.0000000000000
\caption{Series coefficients for the 
triplet gap $\Delta/J'$ at ${\bf k}=0$ 
and staggered susceptibility $\chi_{AF}$ obtained from the 
plaquette expansion  without diagonal bonds (PE1), 
and with  diagonal bonds (PE2) for $J/J'=1.25,1.4$. Series coefficients of $\lambda^i$
 are listed.
}
 \label{tab3}
\begin{tabular}{r|rrrr}
\multicolumn{1}{c|}{} 
&\multicolumn{2}{c}{PE1}&\multicolumn{2}{c}{PE2} \\
\multicolumn{1}{c|}{$i$}&\multicolumn{1}{c}{$J/J'=1.25$} &\multicolumn{1}{c}{$J/J'=1.4$}
                        &\multicolumn{1}{c}{$J/J'=1.25$} &\multicolumn{1}{c}{$J/J'=1.4$}   \\
\tableline
\multicolumn{5}{c}{triplet gap $\Delta/J'$ at ${\bf k}=0$ } \\
  0 & \dec  1.000000000000                 & \dec  1.000000000000                 & \dec  1.000000000000                 & \dec  1.000000000000       \\
  1 & \dec -5.000000000000$\times 10^{-1}$ & \dec -4.000000000000$\times 10^{-1}$ & \dec -9.166666666667$\times 10^{-1}$ & \dec -8.666666666667$\times 10^{-1}$ \\
  2 & \dec -1.949508101852$\times 10^{-1}$ & \dec -1.558796296296$\times 10^{-1}$ & \dec  5.848101551222$\times 10^{-4}$ & \dec -1.753739316239$\times 10^{-2}$ \\
  3 & \dec -1.215547789271$\times 10^{-2}$ & \dec  6.279063786008$\times 10^{-3}$ & \dec  3.919269654832$\times 10^{-1}$ & \dec  7.809106177672$\times 10^{-1}$ \\
  4 & \dec -4.262073088443$\times 10^{-2}$ & \dec -3.063595079003$\times 10^{-2}$ & \dec -1.209759818705                 & \dec -2.966617086045       \\
  5 & \dec -1.169575496893$\times 10^{-2}$ & \dec -4.837854449338$\times 10^{-3}$ & \dec  2.808227188528                 & \dec  1.165893864520$\times 10^{1}$ \\
  6 & \dec -2.087090156347$\times 10^{-2}$ & \dec -1.752776610250$\times 10^{-2}$ & \dec -7.118402535782                 & \dec -5.396727727606$\times 10^{1}$ \\
  7 & \dec -9.640627326157$\times 10^{-3}$ & \dec -8.067626260181$\times 10^{-3}$ &                                      &                                     \\
\tableline
\multicolumn{5}{c}{staggered susceptibility $\chi_{AF}$} \\
  0 &\dec  1.333333333333      &\dec  1.333333333333      &\dec  1.333333333333     &\dec  1.333333333333      \\
  1 &\dec  1.333333333333      &\dec  1.066666666667      &\dec  2.444444444444     &\dec  2.311111111111      \\
  2 &\dec  1.250353652263      &\dec  7.572093621399$\times 10^{-1}$ &\dec  2.323016302459      &\dec  1.684709846149    \\
  3 &\dec  1.201753927683      &\dec  5.539876821845$\times 10^{-1}$ &\dec  2.280972931735      &\dec  1.010069843149    \\
  4 &\dec  1.113460517630      &\dec  3.809246267771$\times 10^{-1}$ &\dec  2.075318140109      &\dec  3.056543491121$\times 10^{-1}$ \\
  5 &\dec  1.044513939707      &\dec  2.747370398592$\times 10^{-1}$ &\dec  2.415891359693      &\dec  6.618315713507$\times 10^{-1}$ \\
  6 &\dec  9.707650534281$\times 10^{-1}$ &\dec  1.964730571491$\times 10^{-1}$ &\dec  2.233378054361      &\dec  4.157664670338$\times 10^{-1}$ \\
\end{tabular}
\end{table}

\begin{table}
\squeezetable
\setdec 0.0000000000000
\caption{Series coefficients for the staggered magnetization $M$ obtained from the 
Ising expansion about N\'eel-order (N\'eel), and the helical-order (helical) expansion
 for $J/J'=1.25,1.4$.
Series coefficients of $\lambda^i$
 are listed.}
 \label{tab2}
\begin{tabular}{r|rrrr}
\multicolumn{1}{c|}{} 
&\multicolumn{2}{c}{helical}&\multicolumn{2}{c}{N\'eel} \\
\multicolumn{1}{c|}{$i$}&\multicolumn{1}{c}{$J/J'=1.25$} &\multicolumn{1}{c}{$J/J'=1.4$}
                        &\multicolumn{1}{c}{$J/J'=1.25$} &\multicolumn{1}{c}{$J/J'=1.4$}   \\
\tableline
  0 & \dec  5.000000000000$\times 10^{-1}$ & \dec  5.000000000000$\times 10^{-1}$ & \dec  5.000000000000$\times 10^{-1}$ & \dec  5.000000000000$\times 10^{-1}$ \\
  1 & \dec  0.000000000000       & \dec  0.000000000000       & \dec  0.000000000000       & \dec  0.000000000000       \\
  2 & \dec -1.976862628853$\times 10^{-1}$ & \dec -1.787535427319$\times 10^{-1}$ & \dec -3.265306122449$\times 10^{-1}$ & \dec -3.906250000000$\times 10^{-1}$ \\
  3 & \dec  8.015745892489$\times 10^{-2}$ & \dec  6.696385209168$\times 10^{-2}$ & \dec  2.332361516035$\times 10^{-1}$ & \dec  3.417968750000$\times 10^{-1}$ \\
  4 & \dec -2.020938451061$\times 10^{-1}$ & \dec -1.703648728139$\times 10^{-1}$ & \dec -6.249672540570$\times 10^{-1}$ & \dec -1.188130158850       \\
  5 & \dec  1.595089057636$\times 10^{-1}$ & \dec  1.186622673353$\times 10^{-1}$ & \dec  1.247039703760       & \dec  2.891792139388       \\
  6 & \dec -2.856270160505$\times 10^{-1}$ & \dec -2.387988493110$\times 10^{-1}$ & \dec -2.564444941065       & \dec -7.392698990572       \\
  7 & \dec  3.509646624490$\times 10^{-1}$ & \dec  2.550333854825$\times 10^{-1}$ & \dec  5.414869562182       & \dec  1.960927054837$\times 10^{1}$ \\
  8 & \dec -6.159923626864$\times 10^{-1}$ & \dec -4.541054989763$\times 10^{-1}$ & \dec -1.300042701432$\times 10^{1}$ & \dec -5.777973112981$\times 10^{1}$ \\
  9 & \dec  9.506010225594$\times 10^{-1}$ & \dec  6.635675648444$\times 10^{-1}$ & \dec  3.100311343644$\times 10^{1}$ & \dec  1.688382921512$\times 10^{2}$ \\
 10 & \dec -1.636489163183       & \dec -1.143031741464       & \dec -7.335660790958$\times 10^{1}$ & \dec -4.957990614738$\times 10^{2}$ \\
 11 & \dec  2.726180882130       & \dec  1.887668026020       & \dec  1.769340004349$\times 10^{2}$ & \dec  1.478774628135$\times 10^{3}$ \\
 12 &                           &                           & \dec -4.347120802018$\times 10^{2}$ & \dec -4.471290404091$\times 10^{3}$ \\
\end{tabular}
\end{table}

\begin{table}
\squeezetable
\setdec 0.00000000
\caption{The results of $\{m/n/l\}$ integrated differential approximants to the series 
 for the ground state $E_0/NJ'$ for $J/J'=1.4$ from the N\'eel ordered Ising expansion and the plaquette 
 expansion without diagonal bonds (PE1). An asterisk denotes a defective
approximant.
}
 \label{tab_e0_ida}
\begin{tabular}{rrrrrrrr}
\multicolumn{1}{c}{n} &\multicolumn{1}{c}{$\{(n-3)/n/l\}$}&\multicolumn{1}{c}{$\{(n-2)/n/l\}$}
&\multicolumn{1}{c}{$\{(n-1)/n/l\}$} &\multicolumn{1}{c}{$\{n/n/l\}$}&\multicolumn{1}{c}{$\{(n+1)/n/l\}$}
&\multicolumn{1}{c}{$\{(n+2)/n/l\}$}&\multicolumn{1}{c}{$\{(n+3)/n/l\}$} \\
\tableline
\multicolumn{8}{c}{N\'eel ordered Ising expansion} \\ 
\multicolumn{8}{c}{$l=0$} \\ 
 n= 3 &            & -0.52170   & -0.55795   & -0.54979   & -0.57974   & -0.55498   & -0.57207    \\
 n= 4 &  *         & -0.55178   & -0.55573   & -0.56195   & -0.56468   & -0.56026   & -0.55991    \\
 n= 5 & -0.56212   & -0.57002   & -0.56907   & -0.56264   & -0.55989   &            &             \\
 n= 6 & -0.56912   &  *         & -0.56121   &            &            &            &             \\
 n= 7 & -0.56405   &            &            &            &            &            &             \\
\multicolumn{8}{c}{$l=1$} \\ 
 n= 3 &            & -0.55300   & -0.55127   & -0.55765   & -0.57441   & -0.56242   & -0.55104    \\
 n= 4 & -0.55139   &    *       & -0.55545   & -0.56380   &     *      & -0.56002   &             \\
 n= 5 & -0.56139   & -0.56960   & -0.56066   & -0.56100   &            &            &             \\
 n= 6 & -0.56503   & -0.56097   &            &            &            &            &             \\
\multicolumn{8}{c}{$l=2$} \\  
 n= 2 &            & -0.56352   &    *       & -0.55148   & -0.55131   & -0.60526   & -0.57327    \\
 n= 3 & -0.55123   & -0.55086   & -0.55252   &    *       & -0.56628   & -0.54598   & -0.55427    \\
 n= 4 & -0.55270   &    *       & -0.55923   & -0.55915   &    *       &            &             \\
 n= 5 & -0.56600   & -0.55917   & -0.56079   &            &            &            &             \\
 n= 6 & -0.56213   &            &            &            &            &            &             \\
\multicolumn{8}{c}{$l=3$} \\  
 n= 1 &            &            & -0.56410   & -0.55914   & -0.55010   & -0.56012   & -0.57599    \\
 n= 2 &            & -0.56089   &    *       & -0.55363   &   *        & -0.56992   &  *          \\
 n= 3 & -0.55104   & -0.55440   & -0.55944   & -0.55918   & -0.56186   & -0.56198   &             \\
 n= 4 &    *       & -0.55918   & -0.55923   & -0.56196   &            &            &             \\
 n= 5 & -0.56143   & -0.56258   &            &            &            &            &             \\
\tableline 
\multicolumn{8}{c}{plaquette expansion without diagonal bond (PE1)} \\ 
\multicolumn{8}{c}{$l=0$} \\
 n= 1 &            &            &            &            & -0.55534   & -0.55281   & -0.55647    \\
 n= 2 &            &            & -0.55246   & -0.55301   & -0.55390   & -0.55473   & -0.55482    \\
 n= 3 &            & -0.55297   &    *       & -0.55492   & -0.55484   &            &             \\
 n= 4 & -0.55425   & -0.55448   & -0.55484   &            &            &            &             \\
 n= 5 &   *        &            &            &            &            &            &             \\
\multicolumn{8}{c}{$l=1$} \\
 n= 1 &            &            &            &  *         & -0.55243   &     *      & -0.55344    \\
 n= 2 &            &            & -0.55316   & -0.55574   & -0.55595   & -0.55499   &             \\
 n= 3 &            &  *         & -0.55587   & -0.55467   &            &            &             \\
 n= 4 & -0.55458   & -0.55513   &            &            &            &            &             \\
\multicolumn{8}{c}{$l=2$} \\                                          
 n= 1 &            &            &  *         & -0.55270   &     *      &   *        &     *       \\
 n= 2 &            & -0.55347   & -0.55481   &   *        & -0.55488   &            &             \\
 n= 3 &   *        & -0.55496   & -0.55487   &            &            &            &             \\
 n= 4 & -0.55492   &            &            &            &            &            &             \\
\end{tabular}
\end{table}

\begin{table}
\squeezetable
\setdec 0.00000000000
\caption{$[n/m]$ Dlog Pad\'e approximants to the series for the triplet gap $\Delta({\bf k})$ 
and antiferromagnetic susceptibility $\chi_{AF}$ for the plaquette expansion without  diagonal bonds (PE1). 
An asterisk denotes a defective
approximant.} \label{tab_pade_pe}
\begin{tabular}{rccccc}
\multicolumn{1}{c}{n} &\multicolumn{1}{c}{$[(n-2)/n]$}&\multicolumn{1}{c}{$[(n-1)/n]$}
&\multicolumn{1}{c}{$[n/n]$} &\multicolumn{1}{c}{$[(n+1)/n]$}&\multicolumn{1}{c}{$[(n+2)/n]$} \\
\multicolumn{1}{c}{} &\multicolumn{1}{c}{pole (residue)} &\multicolumn{1}{c}{pole (residue)} 
&\multicolumn{1}{c}{pole (residue)} &\multicolumn{1}{c}{pole (residue)} &\multicolumn{1}{c}{pole (residue)} \\
\tableline
\multicolumn{6}{c}{$\Delta(0,0)$ for $J/J'=1.25$} \\
 n= 1 &                   &                        & \dec 1.4098(1.2718)  & \dec 0.8592(0.2879) & \dec 1.2001(1.0957) \\
 n= 2 &                   & \dec 1.0804(0.6523)     & \dec 1.0530(0.6018) & \dec 1.0305(0.5502) & \dec 1.0181(0.5170) \\
 n= 3 & \dec 1.0555(0.6075) & \dec 0.9247(0.2175)$^*$ & \dec 1.0074(0.4801) &                   &                   \\
 n= 4 & \dec 1.0089(0.4855) &                        &                   &                   &                   \\
\tableline
\multicolumn{6}{c}{$\Delta(0,0)$ for $J/J'=1.4$} \\                                 
 n= 1 &                   &                        & \dec 2.0315(1.9470)  & \dec 0.8107(0.1237) & \dec 1.5546(1.6732) \\
 n= 2 &                   & \dec 1.2949(0.6512)     & \dec 1.2016(0.5172) & \dec 1.1075(0.3712) & \dec 1.0273(0.2532) \\
 n= 3 & \dec 1.2181(0.5462) & \dec 0.7442(0.0227)$^*$ & \dec 0.9352(0.1267) &                   &                   \\        
 n= 4 & \dec 0.9454(0.1386) &                        &                   &                   &                   \\                  
\tableline
\multicolumn{6}{c}{$\Delta(\pi/4, 3\pi/16)$ for $J/J'=1.4$} \\                                 
 n= 1 &                      &                     &\dec  1.6055(1.4266)  &\dec  0.638(0.0896) &\dec  1.4385(2.3132) \\
 n= 2 &      $*$             &\dec  0.9883(0.4300) &\dec  0.9889(0.4307)  &\dec  0.985(0.4240) &\dec  1.0203(0.5055) \\
 n= 3 &\dec  0.9889(0.4307)  &\dec  0.9884(0.4301) &\dec  0.9885(0.4302)  &                      &                     \\
 n= 4 &\dec  0.9885(0.4302)  &                     &                      &                      &                     \\
\tableline
\multicolumn{6}{c}{$\chi_{AF}$ for $J/J'=1.25$} \\
 n= 1 &                    &                    & \dec 0.9830(-0.8461) & \dec 1.2245(-1.6351) & \dec 0.9962(-0.7165) \\
 n= 2 & \dec 1.0003(-0.8898) & \dec 1.0739(-1.0477) & \dec 1.1011(-1.1273) & \dec 1.0825(-1.0510) &                    \\  
 n= 3 & \dec 1.1033(-1.1367) & \dec 1.0910(-1.0902) &                    &                    &                    \\          
\end{tabular}                                                        
\end{table}

\end{document}